# An Artificial Intelligence Outlook for Colorectal Cancer Screening

Panagiotis Katrakazas
Zelus P.C.
Athens, Greece
ORCID ID: 0000-0001-7433-786X

Aristotelis Ballas
Dept. of Informatics and Telematics
Harokopio University of Athens
Athens, Greece
ORCID ID: 0000-0003-1683-8433

Marco Anisetti
Dipartimento di Informatica (DI)
Università degli Studi di Milano
Milan, Italy
ORCID ID: 0000-0002-5438-9467

Ilias Spais
Zelus P.C.
Athens, Greece
e-mail: ilias.spais@zelus.gr

*Abstract*—Colorectal cancer is the third most common tumor in men and the second in women, accounting for 10% of all tumors worldwide. It ranks second in cancer-related deaths with 9.4%, following lung cancer. The decrease in mortality rate documented over the last 20 years has shown signs of slowing down since 2017, necessitating concentrated actions on specific measures that have exhibited considerable potential. As such, the technical foundation and research evidence for blood-derived protein markers have been set, pending comparative validation, clinical implementation and integration into an artificial intelligence enabled decision support framework that also considers knowledge on risk factors. The current paper aspires to constitute the driving force for creating change in colorectal cancer screening by reviewing existing medical practices through accessible and non-invasive risk estimation, employing a straightforward artificial intelligence outlook.

## I. Introduction

Colorectal cancer is the third most common tumor in men and the second in women, accounting for 10% of all tumors worldwide. It ranks second in cancer-related deaths with 9.4%, following lung cancer. About 1.9 million new cases were diagnosed in 2020, translating into 0.9 million deaths, while incidence is projected to rise significantly over the next decade, with 3.2 million new diagnoses annually by 2040. In affected European Union (EU) individuals, 5-year survival ranges from 28.5% to 57% in men and from 30.9% to 60% in women, with pooled estimations in 23 countries of 46.8% and 48.4% respectively [1], [2]. Moreover, colorectal cancer is among the five most likely to metastasize cancers. Upon initial diagnosis, 22% of cases are metastatic, while about 70% of patients will eventually develop metastatic relapse [3].

In the colorectal cancer treatment domain, options include endoscopic and surgical excision, radiotherapy, immunotherapy, palliative chemotherapy, targeted therapy, extensive surgery and local ablative therapies for metastases [1]. Meanwhile, screening methods consisting of endoscopic tests (e.g., colonoscopy) and non-invasive alternatives such as the fecal occult blood test (FOBT) have been put into action. Applied pathways have successfully inhibited cancer progression [4] contributing to decreased mortality rates through 2017. Moreover, some EU countries have adopted population-based screening programs over the last 15 years, seeking to halt incidence and mortality rates. In this regard, studies have compared mortality rates for symptom-detected vs screening-detected colorectal cancer, stating the considerable impact of screening via quantified reduction estimates surpassing 30% for screening-based detections [5]. Particularly, 5-year survival rate can reach 90% for stage I diagnosis, being less than 15% for advanced stages [6]. Therefore, routine screening is key for reducing mortality and declining incidence rates, since colorectal cancer is now considered a highly preventable disease with a considerably wide temporal development window [7]. Namely, the transitional path from normal mucosa to pre-malignant growth and then to malignant lesion might spread over 15 to 20 years, with scientists seeking means for earlier, cost-effective and less taxing detection of pre-malignant states.

In determining the colorectal cancer risk status, factors such as age, body mass index, diet, smoking habits and family history have been pinpointed by researchers and clinicians alike [8]. Namely, age, sex and family history have been integrated into practice as flagships on risk stratification [2], although evidence on the complete risk factor set has not been analyzed in the context of a detailed assessment. Indicatively, Western registry data show an increased incidence in the age group of 40-44, considerably lower than the 50-year threshold [9]. This tendency is attributed to modern lifestyle alterations, although more assays are required as to the corresponding effects. Overall, despite the long-assumed colorectal cancer preventability based on modifiable risk factors, awareness and knowledge exploitation remain extremely low.

Moreover, despite the available arsenal of screening practices, citizen participation is hindered due to suboptimal performance and invasive or overall taxing nature besetting these methods [10]. EU reports indicate participation rates of 14%, a rather disappointing number compared to the >60% rate for breast cancer screening programs [11]. Poor screening outreach is augmented by limited penetration of Council recommendations into clinical practice. By 2019, only three member states had adopted population-based screening targeting all citizens at risk, albeit purely based on age thresholds of 50-74 years. Over the last three years, other states have also launched population-based screening or regional programs. Still, there is little progress on standardized programs that unite knowledge on colorectal cancer towards EU-wide regulations. Overall, taxing procedures, citizen reluctance, poor awareness and screening accessibility are hindering participation, forcing researchers into the survey of accessible, non-invasive biomarkers that bear the potential to render cancer screening less burdensome and more accessible to citizens. In this vein, liquid biopsy appears to be a promising new tool on non-invasive, quick and safe assessment [6]. However, lack of operating protocols, reproducibility issues and cost-effectiveness discrepancies are impeding clinical implementation. Among all liquid biopsy products, blood-derived proteins seem to constitute the most cost-effective solution regarding resources, sensitivity and research maturity. On this premise, a vast protein pool has been tested, albeit evidence lacks comparative validation, perplexing standardization margins. The current research highlights (a) the existing practices in colorectal screening methodology along with a comparison matrix of their benefits and limitations, along with (b) a risk factor analysis that can be feasibly exploited for population-based screening and sustainably covered by health insurance bodies, via an artificial intelligence (AI)-enabled suggestion

## II. STANDARD OF CARE ON COLORECTAL SCREENING

Early cancer detection through effective screening constitutes the hallmark for providing patients and clinicians with the best possible "fighting chance", more so for colorectal cancer, a highly preventable cancer with long (>10 years) pre-malignant underlying processes. Specifically, the colorectal cancer field has been provided with a variety of screening tools [12], [13], which can be divided into two main groups (Table I). Each method bears specific benefits and limitations with regard to invasiveness, sensitivity, specificity, preparation, sampling process, risks and costs (Table 2).

TABLE I. COLORECTAL CANCER SCREENING TESTS

| Group | Description | Tests | Assessment |
|---|---|---|---|
| Stool-Based | They are simpler and less invasive than structural tests, however they require frequent reassessment | Fecal immunochemical test (FIT) | Every year |
| | | guaiac-Based Fecal Occult Blood Test (gFOBT) | |
| | | Multi-targeted stool DNA test (mt-sDNA) | Every three years |
| Structural | They scan colon and rectum structure for abnormal areas. They are invasive tests conducted either with a scope inside the rectum or via special imaging. | Flexible Sigmoidoscopy (FSIG) | Every five years |
| | | Computed Tomography (CT) Colonography | |
| | | Colonoscopy | Every ten years |

TABLE II. COLORECTAL CANCER SCREENING TESTS

| Test | Benefits | Limitations |
|---|---|---|
| FIT | o *No direct risk to the colon*<br>o *No bowel preparation*<br>o *No pre-test diet or medication changes*<br>o *At-home sampling*<br>o *Fairly inexpensive* | − Can miss many polyps and some cancers<br>− Potential false-positive results<br>− Needs annual reassessment<br>− Requires colonoscopy if abnormal |
| gFOBT | o *No direct risk to the colon*<br>o *No bowel preparation*<br>o *At-home sampling*<br>o *Inexpensive* | − Can miss many polyps and some cancers<br>− Potential false-positive results<br>− Pre-test diet changes and medication changes<br>− Needs annual reassessment<br>− Requires colonoscopy if abnormal |
| mt-sDNA | o *No direct risk to the colon*<br>o *No bowel preparation*<br>o *No pre-test diet or medication changes*<br>o *At-home sampling* | − Can miss many polyps and some cancers<br>− Potential false-positive results<br>− Needs reassessment every three years<br>− Requires colonoscopy if abnormal |
| FSIG | o *Fairly quick and safe*<br>o *Usually doesn't require full bowel prep*<br>o *Sedation usually not used*<br>o *Does not require a specialist*<br>o *Performed every five years* | − Not widely used as a screening test<br>− Examines only about 1/3 of the colon<br>− Can miss small polyps<br>− May cause some discomfort<br>− Small risk of bleeding, infection or bowel tear<br>− Requires colonoscopy if abnormal |
| CT Colonography | o *Fairly quick and safe*<br>o *Can usually examine the entire colon*<br>o *Performed every five years*<br>o *No sedation needed* | − Can miss small polyps<br>− Full bowel prep needed<br>− Potential false-positive results<br>− Exposure to radiation<br>− Inability to remove polyps during testing<br>− Requires colonoscopy if abnormal |
| Colonoscopy | o *Can usually examine the entire colon*<br>o *Can include biopsy and polyp removal*<br>o *Performed every ten years* | − Can miss small polyps<br>− Requires full bowel preparation & (usually) sedation<br>− More expensive than other tests<br>− Small risk of bleeding, bowel tears or infection |

Evidently, in the context of colorectal cancer screening, the medical community is faced with the paradox of a dedicated screening procedure pool that is heavily under-utilized by citizens. Although colonoscopy is recommended as the current gold standard, the long preparation and recovery procedures (~24 hours) as well as the potential adverse effects (colon tears, diverticulitis, abdominal pain, and risks in cardiovascular conditions) induce heavy reluctance for individuals to include colonoscopy in their regular check-up routine. Moreover, all existing Standard-of-Care (SoC) screening tests can miss small polyps, thus failing to detect potential pre-malignant lesions.

## III. STATE-OF-THE-ART ON RISK FACTOR ANALYSIS

Numerous studies have investigated the association of colorectal cancer incidence with demographic, behavioral, and environmental risk factors including age, sex and lifestyle. For instance, a 25% higher incidence has been documented for males, varying among countries [2], while even race has been highlighted as a noteworthy parameter [14]. Overall, age comprises the main factor assessed by current guidelines, formulating at-risk groups for recommended screening. Guidelines suggest screening after 50 years, with healthy citizens advised to pursue regular testing through the age of 75. For people aged 76-85, guidelines are based on overall health and prior history, while beyond this range there are no strict recommendations. However, clinical practice has shown that these thresholds are gradually decreasing, a fact under investigation by the medical community. This status creates a pressing need for alternative practices beyond age-only recommendation onset. Currently, the only such criteria correspond to medical history, family history or symptom manifestation. Indicatively, a meta-analysis of observational studies found that having at least one affected first-degree relative (parents, siblings, or children) increased the risk of colorectal cancer by 2.2-fold and having at least two affected first-degree relatives increased the risk of colorectal cancer by almost 4-fold [15]. Moreover, patients with persistent Inflammatory Bowel Disease (IBD) are twice as likely to acquire colorectal cancer. Inflammation causes aberrant growth cytokines to be released, as well as increased blood flow, metabolic free radicals, and other variables that contribute to carcinogenesis [16]. Regarding symptoms, colorectal cancer may not cause any right away, thus constituting a particularly threatening cancer that necessitates early reliable screening. Alarming symptoms include rectal bleeding or changes in bowel habits such as diarrhea, constipation, or stool narrowing. In any other case, an individual is assigned with a "higher-than-average" risk status under the presence of one (or more) of the following:

- Personal history or family history of colorectal cancer or certain types of polyps
- Personal history of inflammatory bowel disease (ulcerative colitis or Crohn's disease)
- Confirmed or suspected hereditary cancer syndrome (2%–5% of all colorectal cancers), such as familial adenomatous polyposis coli and its variants (1%), Lynch-associated syndromes (hereditary non-polyposis colon cancer) (2%–4%), Turcot, Peutz–Jeghers and MUTYH-associated polyposis syndrome
- Personal history of radiation treatment on the abdomen or the pelvic area for a prior cancer

The above elements comprise unalterable facts regarding early detection or even cancer prevention. On the other hand, several lifestyle-related factors have been identified, which are modifiable through suitable behavioral screening and personalized interventions. In fact, the links of diet, weight and exercise to colorectal cancer risk are some of the strongest among all cancer types. For example, being overweight raises both the incidence and the mortality risk for colorectal cancer in both men and women, but the link seems to be stronger in men. Body-mass index (BMI) and waist circumference (WC) are well-established risk factors for colorectal cancer, as evidenced by epidemiological research employing a variety of anthropometric measurements [17]. By extension, physical activity seems to constitute a key factor with evidence not favoring a sedentary lifestyle, as is the case for dietary habits. Namely, a diet that is high in red meats (such as beef, pork, lamb, or liver) and processed meats (like hot dogs and some luncheon meats) is assumed to raise colorectal cancer incidence risk. Even cooking-related processes seem to play a part, with very high temperatures (frying, boiling, or grilling) generating chemicals that might raise the associated risk. Similarly, lifestyle habits like smoking or alcohol consumption are linked to colorectal cancer incidence [18]. Although smoking is a well-known factor for lung cancer, research has displayed association with additional malignancies. Similarly, heavy alcohol use may cause a number of significant health-related outcomes, with colorectal cancer bearing a connection. By extension, all these modifiable factors have the potential to be addressed via lifestyle interventions promoting healthy behaviors including physical activity, BMI control, appropriate eating/cooking habits and refraining from smoking or excessive alcohol. On the whole, although a large pool of risk factors has been assumed to correlate with colorectal cancer (among other malignancies as well), the underlying regulatory processes remain largely unknown. The key to quantifying the corresponding transition mechanisms might lie into specific biomarkers that are extracted in a minimally invasive and cost-effective manner. Indicatively, it is a well-known fact that dietary habits are translated into alterations in routine blood biomarkers, such as the level of vitamin D in the blood, directly associated with eating patterns [19]. However, such phenotype manifestations have not been analyzed within a colorectal cancer-centered protocol. A handful of studies have assessed some indicative biomarkers, albeit without proposing an interpretation model for behavior effects on biomarker alteration and colorectal cancer incidence [20], [21].

## IV. THE POTENTIAL OF AI IN COLORECTAL CANCER SCREENING

Generally, outcome prediction has been the hallmark regarding AI utilization on cancer. Indicatively, AI models have attempted to predict 5-year colorectal cancer recurrence risk, outperforming grading and/or staging evaluation by expert pathologists. As far as screening is concerned, AI has been implemented for automated decision making in combination with various studied screening procedures [7]. Relying on the duration of the gradual transition path from normal mucosa to a premalignant growth and then to a malignant lesion spanning over 15-20 years, AI can detect suspect changes corresponding to abnormal tissue, which may be indicative of either a premalignant precursor lesion or an early-stage tumor. In particular, a high adenoma detection rate (ADR) has been validated as inversely correlated with adenoma miss rate and the risk of post-colonoscopy colorectal cancer), with every 1% increase in ADR corresponding to a 3% reduction in colorectal cancer development risk and 5% reduction in fatal colorectal cancer incidence [22]. However, ADRs may range from 7% to 53% among different endoscopists, creating the demand for objective and reproducible assessment towards attaining a robust ADR in clinical settings. In this regard, convolutional neural networks (CNNs) have been found to accurately detect and localize premalignant lesions on imaging data. In a prospective randomized trial under controlled conditions [23] with conventional vs computer-aided colonoscopy detection, ADR was significantly increased in favor of computer-aided detection (CAD). Similarly, virtual colonoscopy paired with machine learning (ML) modules was able to distinguish benign and precancerous colorectal polyps in an average-risk asymptomatic colorectal cancer screening samples with a sensitivity of 82% and specificity of 85%.

In the field of liquid biopsy, AI impact has been examined with the use of supervised learning methods such as support vector machines (SVMs). Hierarchical classification has shown feasible applicability, with an initial classification level filtering out non- colorectal cancer samples and a subsequent level differentiating between non-malignant lesions and a "no findings" class [24]. Additionally, blood biomarkers have been considered for use, alongside risk factors (e.g., demographic), within an AI-based framework for estimating cancer risks. These biomarkers were not limited in complex resource-demanding analytics, but included standard blood test markers as well, such as red cell distribution width (RDW) and anemia findings [25], [26] . Further knowledge can be drawn from electronic health records (EHR), thus incorporating screening in primary healthcare settings.

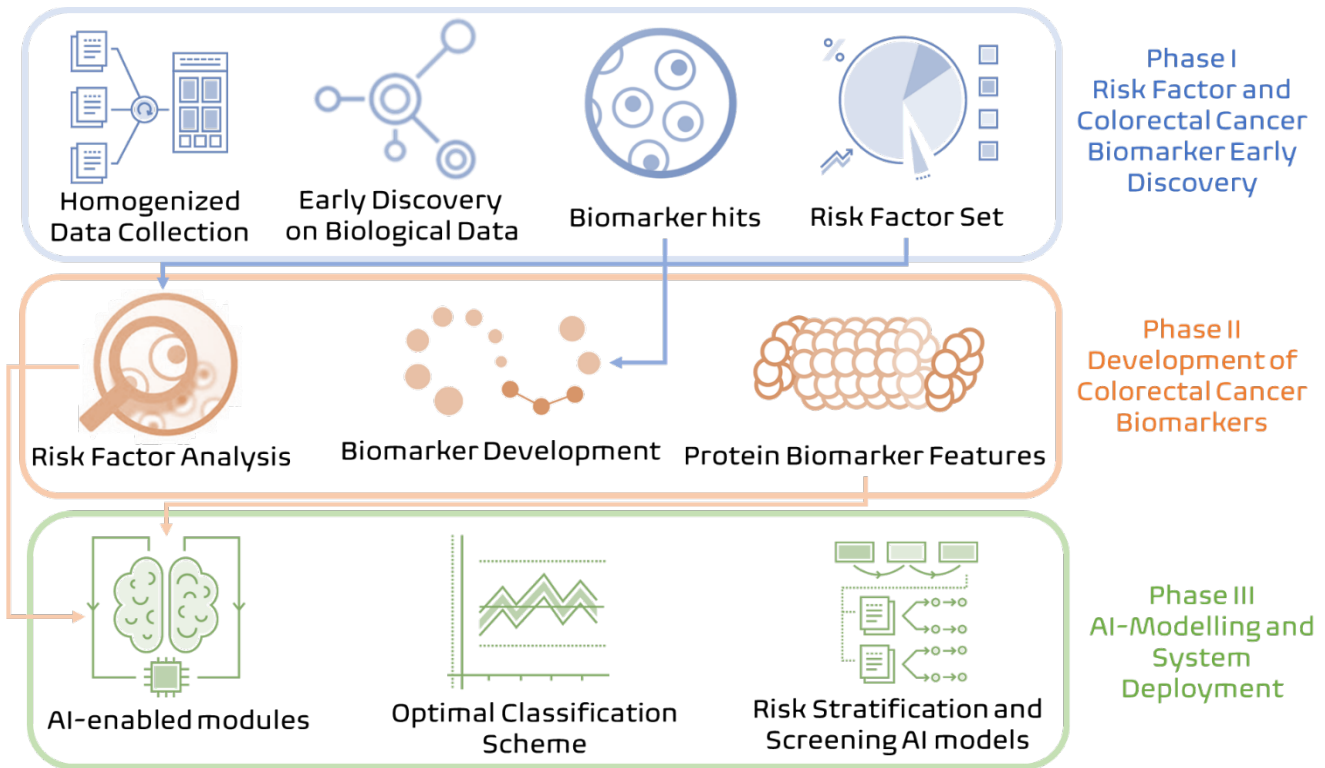

*Figure 1 Suggested Methodological Architecture of the AI-based solution*

Notably, combining such models with gFOBT seems to contribute to more than 2-fold increase in cancer detection capability for retrospective data cohorts. Based on this information pool, decision trees can be constructed for producing risk stratification scores and proposing clinical pathways. Finally, with regard to the Cancer of Unknown Primary (CUP) case, AI has been used with known metastatic colorectal cancer cases as training data, attempting subsequent testing and validation on both metastatic colorectal cancer and CUP data [27].

AI-powered virtual assistants can also provide personalized healthcare services and improve communication between patients and care providers [7]. By this principle, AI-based mobile applications such as the Colorectal Cancer Awareness Application (ColorApp®) [28] have sought to foster community education and participation in screening programs, achieving a score of 72 in the System Usability Scale Questionnaire for the Assessment of Mobile Apps. However, there is a lack of regional adaptations and performance surveys beyond usability scores.

In general, despite the above AI advances in colorectal cancer risk and progression assessment, the medical community is still skeptical and reluctant in trusting the outcomes of machine learning methods. This is mainly due to the depth of most neural network approaches, which are regarded as "black boxes", along with their confusing architecture. Furthermore, the majority of deep learning models fail to generalize adequately on previously 'unseen' data [29]. Since medical practices and methods are often altered and updated, AI models are expected to maintain their performance when evaluated on data that are sampled from a similar but different distribution than that of the model's training data (e.g., datasets across hospitals or signals captured by different devices). To this end, several Domain Generalization (DG) [30]–[32] methods have been proposed. DG methods attempt to push AI models to extract invariant features present in the data and therefore be less affected by the potential distributional shift in different datasets. In addition, explainable artificial intelligence (XAI) is gradually becoming a prerequisite by clinicians and policy makers, seeking to instill accountability and medical transparency into AI-assisted decisions for launching trustworthy clinical.

V. SUGGESTION OF AN AI-BASED SOLUTION

Under these suggestions, AI can be exploited in the context of assessing an extended risk factor pool compared to current limited clinical consideration and by creating a clinical decision support for stratifying high-risk cases as an output from a front-line screening tool to the main evaluation phase. Figure 1 depicts the suggested approach.

Organization of knowledge into a standardized format, can generate a homogenized dataset. This will be utilized where risk factor investigation will take place, producing a holistic colorectal cancer risk factor set and a set of biomarker 'hits' to be further evaluated during a subsequent analytics step. This phase will in turn generate protein features for training and evaluating an AI model. In the latter phase, risk factor and protein features will undergo an AI-based analysis for generating a proposed subset of discriminative protein biomarkers as well as a full stratification model utilizing a hybrid approach based on the combination of biomarker and risk factor phenotypes.

*A. Phase I: Risk Factor and Colorectal Cancer Biomarker Early Discovery*

Demographic, behavioral, environmental, medical and history factors will be filtered from available data based on expert knowledge and existing research for determining the set that will be used during Phase III in the context of AI model training. In cases where retrospective data are deemed insufficient for AI training regarding individual parameters, those will be incorporated in the subsequent protocol, pending investigation and verification of their role based on prospective data. Regarding colorectal cancer biomarkers from blood samples, the necessary activities following the traditional biomarker discovery pipeline can be conducted by clinical entities, from early discovery and verification to clinical validation. Phase I corresponds to the early discovery phase, involving the identification of "hits" by searching thousands of proteins in limited number of samples that are present in blood serum/plasma using Proximity Extension Assay (PEA) by Olink, utilizing Next Generation Sequencing (NGS) methods to measure the concentration of thousands of human plasma proteins from small sample volumes [27]. PEA uses matched pairs of antibodies attached to unique DNA nucleotides for each protein to produce NGS readouts for the parallel quantification of up to 3000 proteins.

NGS readouts will be analyzed using statistical methods to identify a projected panel of ~30 protein markers that are representative of colorectal cancer incidence, showing a significant difference between healthy and diseased blood samples. However, due to the complex nature of cancer and the lack of mechanistic insight of proteome-wide or large-scale discovery screens, most of the identified biomarkers may still have poor diagnostic performance and clinical outcome. On that account, early discovery will be followed by main biomarker development in Phase II, exploiting an extended pool of retrospective as well as prospective data. Phase II will provide the complete list of colorectal risk factors to be included in the AI modelling process for risk stratification, behavioral monitoring, and interventions.

*B. Phase II: Development of Colorectal Cancer Biomarkers*

As stated above, due to the complex nature of cancer and the lack of mechanistic insight of proteome-wide or large-scale discovery screens, the majority of identified biomarkers can still have poor diagnostic performance and clinical outcome. This is evident by the limited translation of candidate biomarkers into clinical diagnostic assays, with less than two overall approvals per year [28]. Thus, a mechanism-based approach using multi-omics data is required to verify the role of the identified serum biomarkers in the complex colorectal cancer mechanism and select the most promising ones for clinical validation [33]. Recent advances in systems-based approaches have led to the development of interdisciplinary methods that utilize multi-omics data and network-based technologies to elucidate the complex disease mechanism [34]. More specifically, transcriptomic and proteomic data from patients' tissue samples are coupled with knowledge bases of bio-molecular interactions to create in-silico models that best explain experimental data and are representative of the disease mechanism(s) [35]. These models are often represented as signaling networks, with nodes being the signaling proteins and edges depicting the directed flow of information in the system.

Using pathway and network analysis methods, these models can be analyzed to identify the affected biological pathways and central protein nodes that are characteristic of the colorectal cancer mechanism. Being a representation of the colorectal cancer mechanism, these models can be applied to verify the connection and importance of the blood biomarkers identified during the early discovery phase. The result is a selection of verified biomarkers, present in blood, with mechanistic insight and improved performance for the early detection of colorectal cancer. Finally, the identified panel of biomarkers can be developed into highly specific and cost-effective multiplex proteomic assays to provide powerful features for an AI-based detection system.

A state-of-the-art method for multiplex diagnostic assay development involves the use of bead-based immunoassays. Multiplex assays utilize the xMAP technology (Luminex Corp) that relies on color-coded microspheres (bead regions) to allow for the simultaneous detection of responses against multiple protein targets from the same sample. Each bead region is coated with an antibody that recognizes and binds to a specific part of the protein. Mixtures of bead regions are used in a sandwich-type Enzyme-Linked Immunosorbent Assay (ELISA) assay to provide relative and absolute quantification of multiple proteins across the various conditions tested. These assays offer high multiplexability, sample throughput, quality of measurements and specificity for measurement of identified biomarkers in blood. Multiplex readouts from the developed assays can then be combined with logic-based and AI-based computational models to select the optimal combination of biomarkers that maximizes their detection and diagnostic performance [36]. In essence, this step constitutes the main proteinic feature selection process for AI modelling of Phase III towards the identification of the biomarkers associated with optimal screening capacity. Outputs of Phase II will equip with protein features the AI analysis aiming to pinpoint the most significant biomarkers bearing biological explainability through the demonstration of their role within colorectal cancer regulatory network paths.

*C. Phase III: AI-Modelling and System Deployment*

Features pinpointed during Phases I & II (risk factors & proteinic markers) will comprise the necessary feature vector for training and testing the AI-enabled modules. Initially, a pool of state-of-the-art AI methodologies, encompassing supervised learning, unsupervised learning and reinforcement learning for operating on retrospective datasets will be deployed. Parametric and non-parametric techniques will be incorporated with network-based techniques also included, maintaining interpretability by individually testing feature subsets. As such, application of multiple frameworks in pursue of the optimal classification scheme, testing both individual algorithms and ensemble schemes (e.g., majority-vote models) within various combinations and architectures can be pursued. Moreover, enhancing interpretability and transparency can be offered for avoiding class exclusion for cases close to the decision boundaries for colorectal cancer risk. This can be achieved by exploration of probabilistic and fuzzy classification, both for directly producing quantified estimates and for providing clinicians with comparative class predictions. This pursue will remain in the context of deep neural networks (DNNs), that bear high performance, although the multitude of layers and lack of understanding on prediction derivation render explainability a challenging task [37]. To address this, dedicated techniques for investigating the models' decisions such as SHapley Additive exPlanations (SHAP) values [38], saliency maps and Gradient-Based Localization [39] shall be deployed. Furthermore, Layer-wise Relevance Propagation (LRP) propagation-based approaches will take into account the internal structure of the neural network propagating the prediction from the output backwards to the input in order to assign relevance scores to the input variables and in compliance with some predefined rules [40]. Moreover, Local Interpretable Model-agnostic Explanations (LIME) will be applied to the DL architecture, explaining the model's response to local perturbations, occluding some of the inputs to investigate the importance of specific variables.

As part of AI modelling, validation will also constitute a major step with regard to the trustworthiness and reliability of the optimal AI model. In this respect, the large pool of retrospective data comprises the required foundation for extensive evaluation of the tested schemes. By utilizing cross-validation architectures and applying weighted multi-class splitting within the available datasets, it possible to extract unbiased estimations, while applying the traditional evaluation metrics (accuracy, specificity, sensitivity, precision etc.) on a class-specific scale. The above techniques will be applied on the extracted features consisting of serum sample protein features, demographic, behavioral and medical features. Classification will be attempted both individually for each feature class (i.e., separating protein data) and employing combinatorial schemes, attempting to link proteinic profiles with routine factors (modifiable or not), aiming for an enhanced model that valorizes routine data alongside protein markers. This will be done with the vision of generating two separate AI models appertaining to real-life use cases. Namely, the risk stratification AI model (RS-AI) will solely assess risk factors and potential risk adjustments based on suggested behavioral interventions. On the other hand, the screening AI model (S-AI) will assess liquid biopsy results – in combination with risk factors – to provide decision support on high-risk individuals that should undergo colonoscopy.

On the whole, the AI system should also by-principle and by-design fulfil the requirements comprising (i) accountability, (ii) privacy and data governance, (iii) societal and environmental wellbeing, (iv) technical robustness and safety, (v) human agency and oversight, (vi) diversity and fairness, and (vii) transparency. In such manner, the front-end interface will provide relevant knowledge to clinicians, medical professionals and citizens regarding risks and guidelines for colorectal cancer, raising awareness and contributing to enhanced prevention through population knowledge and evidence-based recommendations.

## VI. Conclusion

Strengthening the collaborative environment and building AI-based bridges between different fields of practice- and theory-driven research in colorectal cancer screening is more than necessary given the cost of the existing healthcare pathway. By delivering an early screening and risk assessment solution via AI, solutions can associate significant advancements on many different levels, especially those related to patient acceptance and knowledge discovery. Embracing of new technologies in the colorectal screening pathway will improve public health, while generating knowledge to increase the quality of life. Thus, screening will become more accessible, offering risk assessment, health monitoring and better disease evaluation. Besides early diagnosis, identification of disease trajectories and relapse, as well as employment of cutting-edge technologies may shed light to the nature of the colorectal cancer biological initiation mechanisms, therefore improving the existing human condition.

While colorectal cancer screening tools have proven to be effective, the medical examinations usually employed present several obstacles. As such, a substantial colorectal cancer miss rate has been often reported, while the fact that certain screening methods (e.g., colonoscopy) usually require hospital/clinical settings puts excessive workload on the medical experts to program appointments and consultation. In this regard, information regarding colorectal cancer development could not only prioritize citizens clinic appointments (e.g., increasing the participation of high-risk colorectal cancer patients), but also allow healthcare professionals to be better positioned in terms of the required screening technique.

Allowing preliminary and evidence-based results, our suggested approach will provide medical experts with detailed information for an early and precise screening of colorectal cancer augmenting clinical decision reliability for preventing unneeded invasive examination and thus potential adverse outcomes. Furthermore, it will improve quality in the clinical actions and patient care, reducing the time spent on routine administrative tasks, while providing easy and faster access to colorectal cancer knowledge staging and intervention planning.

The knowledge generated by targeting subject-specific quantification of precise biomarkers determining divergencies and tolerance constraints on colorectal cancer determinants via Phases I & II of the methodology presented here, will be assessed based on cutting-edge AI tools (Phase III). This way, only the clinical validation will be left to be performed via medium or large-scale pilots. Thus, it will grant personalized staging and early protopathic/recurrence prediction, reducing human resources costs by unburdening the healthcare system of nonessential hospital/clinical visits.

## Acknowledgment

This project has received funding from the European Union's Horizon 2020 research and innovation Programme under Grant Agreement No. 875351.